\documentstyle[titlepage,12pt]{article} 
\begin{document}

\title{Cosmologies with Two-Dimensional Inhomogeneity}

\author{A. Feinstein, J. Ib\'a\~nez and Ruth Lazkoz\\
Dpto. F\'{\i}sica Te\'orica, Universidad del Pa\'{\i}s Vasco, Bilbao,
Spain.\\ PACS numbers : 04.20.Jb, 98.90.Hw, 98.90.Cq  } \maketitle

\begin{abstract}

We present a new generating algorithm to construct exact non static solutions
of
the  Einstein field equations with two-dimensional inhomogeneity. Infinite
dimensional families of $G_1$ inhomogeneous solutions with a self
interacting scalar field, or alternatively with perfect fluid, can be
constructed using this algorithm. Some families of solutions and the
applications of the algorithm are discussed.

\end{abstract}

Recently there has been a considerable interest in scalar field cosmologies.
This interest is related to the studies of inflationary models where the scalar
field (inflaton) acts as the main driving force for inflation. On the other
hand, one of the most important issues related to the inflationary scenario
is how sensitive are the inflationary models to the change in the initial
conditions \cite{og}. The dependence of the inflationary scenario on the
initial
conditions  is usually studied either numerically or perturbatively, under some
simplifying assumptions, or analytically, within the framework of exact
solutions using cosmological models with a high degree of symmetry. Until now,
it was only possible to study exact scalar field cosmological models with one
dimensional inhomogeneity \cite{fi}.

The main purpose of
this letter is to make a step further in the study of exact
inhomogeneous cosmologies by presenting a  new generating algorithm to obtain
solutions with a single isometry. The algorithm allows one to obtain
spacetimes with homogeneity broken in two directions from already known ones
with one dimensional inhomogeneity.  One
starts with a vacuum orthogonal
$G_2$ cosmology which is well studied and some general solutions for certain
topologies are known (for a recent review see Ref. \cite{bgm}.) Then, by using
a
known algorithm the solution is transformed into a massless scalar field
$G_2$ cosmology. The second step involves a conformal transformation which
breaks the
$G_2$ symmetry leaving one with the solution  with the $G_1$ symmetry but with
a
self interacting scalar field with a Liouville type of potential.

Our algorithm generalizes a recently proposed solution generating
technique developed by Fonarev \cite{f}, which is only valid for static
metrics.

To this end we consider a vacuum $G_2$ cosmology with the following line
element
\begin{equation}
ds^2 =  e^{f(t,z)} (-dt^2+dz^2)+G(t,z) \left( e^{p(t,z)} dx^2+e^{-p(t,z)}
dy^2\right) \, .
\end{equation}
If \(p_v, G_v\) and \(f_v\) represent a vacuum solution of the Einstein field
equations then, by using a simplified version of the Charach-Malin \cite{cm}
algorithm the following solution
\begin{eqnarray}
\bar{p} & = & B\, p_v+C\,\log G_v \nonumber \\
\bar{f} & = & f_v+E\, p_v+F\, \log G_v \\
\bar{\phi} & = & A\, p_v+D\, \log G_v \nonumber
\end{eqnarray}
will be a solution of Einstein-masless scalar field equations, where the scalar
field \(\bar{\phi}\) is minimally coupled to gravity  and its energy
momentum tensor is given by
\begin{equation}
T_{\alpha \beta}=\bar{\phi}_{,\alpha} \bar{\phi}_{, \beta} -
\frac{1}{2}\,
\bar{g}_{\alpha \beta} \,\,\bar{\phi}_{, \gamma} \bar{\phi}^{,
\gamma}\, .
\end{equation}

The constants \(A, B, C, D, E\) and \(F\) are subject to the
following constraints
\begin{eqnarray}
B\,C-E+2\,A\,D &=& 0 \nonumber \\
C^2+2\,D^2-2\,F &=& 0 \\
B^2+2\, A^2 &=& 1 \nonumber
\end{eqnarray}

\noindent
{\em Theorem}: Given a minimally coupled massless scalar field solution
\(\bar{g}_{ab}\) described by the equations (1)-(3), a new solution with
minimally coupled scalar field with exponential potential can be built by the
following conformal transformation
\begin{equation}
g_{ab}=\Omega(x)^2\,\bar{g}_{ab}\, .
\end{equation}
The new scalar field \(\phi\) will be given by
\begin{equation}
\phi=\bar{\phi}+k\,\log \Omega(x)
\end{equation}
and the potential will take the form
\begin{equation}
V(\phi)=V_0\,e^{-k\phi} \, .
\end{equation}
For \(k^2=2\) the conformal factor \(\Omega\) is given by
\(\Omega=\exp (\alpha x)\) and \(V_0=2\alpha\).
For any other value of \(k\) the conformal factor is given by
\begin{equation}
\Omega=(\alpha x)^{\frac{2}{k^2-2}}, \;\; V_0=\frac{12-2k^2}{(k^2-2)^2}\,
\alpha^2
\end{equation}

The constants \(A, B, C, F\) and \(E\) can be expressed in terms of the
constants
\(D\) and \(k\) which are related with the dynamical and potential parts of the
scalar field respectively:
\begin{eqnarray}
A &=&\pm \frac{1}{\sqrt{k^2+2}} \nonumber \\
B &=&\pm \frac{k}{\sqrt{k^2+2}} \nonumber \\
C &=& D\, k-1 \nonumber \\
F &=& \frac{1}{2}\,(Dk-1)^2+D^2 \nonumber \\
E &=&\pm \frac{1}{\sqrt{k^2+2}}\,\left(k(Dk-1)+2D\right)
\end{eqnarray}

We will not prove the theorem here, but  the results can be verified in
straightforward manner by direct substitution  into the Einstein field
equations.

Before going any further several remarks are in order. This theorem gives one a
simple way to reduce the symmetry of the problem. Generically, one reduces the
symmetry   at the cost of introducing a potential term. However in the special
case \(k^2=6\) in the expression (8) one still obtains solutions with a
massless
scalar field (or stiff fluid.) One does not have to start with the
$G_2$ symmetry but with any larger group which contains the $G_2$. For example,
one can start with $G_3$ Bianchi I model and obtain from it the $G_2$
inhomogeneous solutions. Also, since we have not specified  the character of
the
transitivity surface area which gradient \(G_\mu  G^{\mu}\) can be timelike,
spacelike or null one can construct models starting from the metrics describing
cylindrical gravitational waves, plane waves, cosmic strings, etc..  The new
obtained metrics can be looked at as perfect fluid spacetimes, if one
wishes, by identifying the scalar field with the velocity potential of the
fluid. Thus this is the first algorithm to obtain perfect fluid solutions with
$G_1$ symmetry.

To demonstrate the working power of the algorithm we present several examples.

\bigskip
\noindent
{\em Bianchi I as a seed}. We start with the $G_3$ Bianchi I vacuum (Kasner)
solution by specifying the line element (1) with the metric functions
\begin{equation}
G_v=t, \;\; p_v=\beta\, \log t,\;\; f_v=\frac{\beta^2-1}{2}\,\log t
\end{equation}
The massless scalar field solution will be described by the metric functions
\begin{equation}
\bar{p}=\left( B\beta+C\right)\log t,\; \bar{f}=\left(\frac{\beta^2-1}{2}+
E\beta+F\right)\log t,
\end{equation}
while the scalar field will be given by
\begin{equation}
\bar{\phi}=\left( A\beta +D\right)\log t,
\end{equation}
where the constants are subject to conditions (4). And finally, the solution
with a self interacting scalar field with an exponential potential will be
described by the expressions (5)-(7), where the conformal factor is
\begin{eqnarray}
\Omega(x) &=& (\alpha x)^{\frac{2}{k^2-2}}\quad\hbox{when}\quad k^2\neq
2\nonumber \\
\Omega(x) &=& e^{\alpha x} \quad\hbox{when}\quad k^2=2
\end{eqnarray}
The new solution represents an inhomogeneous $G_2$ cosmology. This solution is
a particular case of more general solutions obtained in \cite{fjl} integrating
directly the Einstein equations.

\bigskip
\noindent
{\em Cosmological gravitational waves as a seed}. Here one starts with the
vacuum  cosmological model describing  propagation of inhomogeneities in
form of gravitational radiation on a spatially flat homogeneous background.
To simplify the discussion we concentrate on a single mode perturbation. One
could easily start from either an infinite-dimensional Gowdy like vacuum
cosmology or inhomogeneous cosmologies filled with monochromatic waves and
pulses \cite{g}. This, however, will not contribute qualitatively to our
discussion. The vacuum solution is given by
\begin{eqnarray}
G_v &=& t \nonumber \\
p_v &=& \beta\,\log t+A_0 \cos (\omega z)\, J_0(\omega t)
\nonumber \\
f_v &=&
\frac{\beta^2-1}{2}\,\log t+\beta\,A_0 \cos (\omega z)\, J_0(\omega
t)+\frac{1}{4}(\omega t)^2\,A_0^2\left[J_0^2(\omega t)+J_1^2(\omega
t)\right]\nonumber \\ & & -\frac{1}{2}\, \omega t\,A_0^2\, \cos ^2(\omega
z)\,J_0(\omega t)\,J_1(\omega t)
\end{eqnarray}

{}From this solution one constructs the minimally massless scalar field
solution
and then using the theorem the exponential potential solution. Since the
construction is straightforward we will not present the explicit form of the
solution. The resulting new solution represents a spacetime with two
dimensional
inhomogeneity. The interpretation of the solution as representing propagation
of gravitational waves is not at all clear now. It would be interesting to
consider if the obtained solution represents the interaction of the
gravitational waves with the scalar field. Work is in progress in this
direction and we hope to be able to present results in the future.

The structure of the solution is much more complicated than that obtained
from the Bianchi I seed. Since the new solutions are always conformally
related with the seeds the asymptotic behaviour at $t=0$ or at late times can
be easily deduced from the asymptotic behaviour of the seed metric. For
example, since usually the $G_2$ cosmologies with massless scalar field at late
times tend to the so called  Doroshkevich-Zeldovich-Novikov
spatially anisotropic universe \cite{dzn} the new solutions with self
 interacting
scalar field will represent an inhomogeneous generalization of DZN
universes at late times.
Near the initial singularity, on the other hand, the structure is more
complicated. It is known that the conformal transformations usually change the
nature of the singularity.

Also, the presence of the potential term for the
scalar field changes the energy balance which can result in violating the
strong
energy condition. It is known that the breaking of the strong energy condition
 is
intimately related with inflation. Note, that in a highly inhomogeneous model
it
is difficult, and sometimes even impossible to see whether the model
inflates. If the cosmological model does not deviate strongly from
homogeneity, say a slightly perturbed model, then the hypersurfaces of
homogeneity perfectly define a preferred observer which can decide as to
whether the model inflates. If the scalar field is not very
inhomogeneous the hypersurfaces of \(\phi=constant\) serve to select again a
preferred observer related to the motion of the perfect fluid which is
defined from the scalar field. Thus, the notion of inflation in regular
spacetimes is well clear. If the spacetime is highly irregular, on the other
hand, neither the scalar field nor  other geometric considerations
help to indicate inflation. In this case one has to use somewhat a weaker way
to
specify inflationary behaviour. For example, one may look at the fulfillment
of the strong energy condition the breaking   of which is a necessary
condition for a model to inflate. Obviously the models which satisfy the strong
energy condition can be considered as the standard ones. We have checked the
fulfillment of the strong energy condition for both models above.

In the inhomogeneous models obtained from the Bianchi I seed (the models
with one dimensional inhomogeneity) the behaviour to respect inflation
 is quite simple. In general, at early
times and for \(k^2<2\) the strong energy condition is fulfilled and the
cosmological model behaves in the standard way. Later the energy condition
breaks down and the model starts to inflate. This also depends on the spatial
coordinate $x$. In the models obtained from the $G_2$ seed the behaviour
presents much more complicated structure. In Fig. 1 we have depicted the
function $3p+\rho$ in terms of coordinates $x$ and $z$ for a constant value of
time $t$. One can easily see that there is a ``wave-like'' pattern along the
$z$
direction, with energy condition broken periodically. As the time passes the
model starts to inflate everywhere, and the ``wave-like'' pattern gets washed
away.

{\em Two solitons as a seed}. One may start with a vacuum cosmological model
describing the propagation and interaction of two strong  gravitational pulses
in a homogeneous background \cite{cv}. The solution for the functions $p_v$and
$G_v$ is given by:
\begin{eqnarray}
G_v &=& t \nonumber \\
p_v &=& \beta \log t-d_1\cosh^{-1} \frac{z+z_1}{t}-d_2\cosh^{-1}
\frac{z+z_2}{t}
\end{eqnarray}
where the constants $z_1$ and $z_2$ can be complex in which case the real
part of the function $p_v$ must be taken. The vacuum seed behaves as follows
\cite{cv}: at early times the model is highly irregular while at late times
the solution tends to the background which is of Kasner type. Applying the
generating algorithm one  obtains a model which will have two
dimensional inhomogeneity at early times while at late times the model will
tend to those obtained from Bianchi I seed.

 In the case when the soliton ``poles'' are 	real ($z_1$ and $z_2$ are
real) it
is well known that the former seed solution represents the interaction
region of the collision of two plane gravitational waves on a flat
background. It is tempting, therefore, to interpret the new solution
obtained after applying the algorithm as a collision of gravitational
waves in the presence of a self interacting scalar field. This solution can
give one some ideas as to how gravitational waves interact in the presence of
a non trivial fluids.

To this end it is important to stress that since there are
a great amount of known
$G_2$ solutions with different physical interpretations, topologies,
singularity
structures etc.
\cite{bgm} the new  algorithm represents a powerful tool to obtain
new families of $G_1$ solutions with two dimensional
inhomogeneities. In certain cases, imposing additional conditions on
the scalar field (the timelike character of its gradient) the new solutions
can be considered as perfect fluid ones.

We have tried to present in this Letter a brief insight on the possibilities of
the application of the new algorithm. We believe that the algorithm presented
here opens a new door into the study of exact solutions of Einstein equations
depending on three variables. The exact solutions obtained with this algorithm
could also serve as test solutions for numerical relativity. The analysis and
 the study of various particular cases are left for future works.

\vspace{1cm}
\noindent
This work is supported by the Spanish Ministry of Education grant (CICYT)
PB93-0507. R.L.'s work is supported by Basque Government fellowship BFI94-094



\begin{thebibliography}{25}

\bibitem[1]{og} K.Olive, Phys. Rep. {\bf 190} (1990) 307. D.S.Goldwirth and
T.Piran, Phys. Rep. {\bf 214} (1992) 224.

\bibitem[2]{fi} J.M.Aguirregabiria, A.Feinstein and J.Ib\'a\~nez, Phys. Rev D
{\bf 48} (1993) 4669.

\bibitem[3]{bgm} W.B.Bonnor, J.B.Griffiths and M.A.H.MacCallum, Gen. Rel. Grav.
{\bf 26} (1994) 687.

\bibitem[4]{f} O.A.Fonarev, {\em Exact Einstein-scalar field solutions for
formation of black holes in a cosmological setting}  Preprint gr-qc 9409020.

\bibitem[5]{cm} Ch.Charach, S.Malin, Phys. Rev. D {\bf 19} (1979) 1058.

\bibitem[6]{fjl} A.Feinstein, J.Ib\'a\~nez and P.Labraga, {\em Scalar field
inhomogeneous cosmologies} submitted for publication.

\bibitem[7]{g} R.Gowdy, Phys. Rev. Lett. {\bf 12} (1971) 60. D.J.Adams,
R.W.Hellings, R.L.Zimmerman, H.Farhoosh, D.I.Levine and Z.Zeldich, Astrophys.
J. {\bf 253} (1982) 1.

\bibitem[8]{dzn} A.G.Doroshkevich, Ya.B.Zeldovich and I.D.Novikov, Sov. Phys.
JETP {\bf 26}, 408 (1968)

\bibitem[9]{cv} B.J.Carr and E.Verdaguer, Phys. Rev. D {\bf 28}
(1983) 2995. J.Ib\'a\~nez and E.Verdaguer, Phys. Rev. Lett. {\bf 51} (1983)
1313. A.Feinstein and Ch.Charach, Class. Quantum Grav. {\bf 3} (1986) 2995.

\end{thebibliography}
\end{document}